# A portable diagnosis model for Keratoconus using a smartphone

Yifan Li[a], Peter Ho[b], Jo Woon Chong [c]

[a] Department of Electrical and Computer Engineering, Texas Tech University, Lubbock, USA,

[b] Lubbock Eye Clinic, Lubbock, Texas 79410, USA

[c] Department of Electrical & Computer Engineering, Sungkyunkwan University, Jongno District, Seoul, South Korea, jwchong@skku.edu

**Abstract**

Keratoconus (KC) is a corneal disorder that results in blurry and distorted vision. Traditional diagnostic tools, while effective, are often bulky, costly, and require professional operation. In this paper, we present a portable and innovative methodology for diagnosing. Our proposed approach first captures the image reflected on the eye's cornea when a smartphone screen generated-Placido disc sheds its light on an eye, then utilizes a two-stage diagnosis for identifying the KC cornea and pinpointing the location of the KC on the cornea. The first stage estimates the height and width of the Placido disc extracted from the captured image to identify whether it has KC. In this KC identification, k-means clustering is implemented to discern statistical characteristics, such as height and width values of extracted Placido discs, from non-KC (control) and KC-affected groups. The second stage involves the creation of a distance matrix, providing a precise localization of KC on the cornea, which is critical for efficient treatment planning. The analysis of these distance matrixes, paired with a logistic regression model and robust statistical analysis, reveals a clear distinction between control and KC groups. The logistic regression model, which classifies small areas on the cornea as either control or KC-affected based on the corresponding inter-disc distances in the distance matrix, reported a classification accuracy of 96.94%, which indicates that we can effectively pinpoint the protrusion caused by KC. We conducted a statistical analysis, yielding a significant p-value less than $10^{-6}$, underscoring the considerable differences in the inter-disc distances between the control and KC groups. Additionally, we augmented the interpretability of our distance matrix by implementing a color scheme, providing valuable insights into the precise bulged location result from KC and the severity of the cornea. This comprehensive, smartphone-based method is expected to detect KC and streamline timely treatment.

**1. Introduction**

Keratoconus (KC) is a corneal disorder causing blurry and distorted vision from thinning and steepening of the cornea, as shown in Figure 1a (Rabinowitz, 1993, 1998). The potential risk factors of KC are corneal thinning and subsequent steepening associated with Down syndrome, atopy, eczema, excessive eye rubbing, and genetic disorders (Gordon-Shaag et al., 2015; Karolak & Gajecka, 2017). Moreover, environmental factors, such as long-lasting ultraviolet light exposure and hot climate in some Asian countries such as India, may result in an increase in the number of KC subjects compared to other parts of the world (Assiri et al., 2005; Nielsen et al., 2007).

The most prevalent method for diagnosing KC is monitoring clinical signs such as Vogt striae, Rizzuti signs, Munson signs, Fleischer rings, and the retinoscopy scissoring reflex by using slit-lamp examination (Jonas et al., 2009). These methods can quickly diagnose moderate to advanced cases of KC. However, the early detection of KC helps to prevent vision loss or the need for corneal transplantation, i.e., timely intervention with corneal cross-linking (CXL) can be performed before the cornea has thinned excessively, and early detection can help avoid complications associated with advanced keratoconus (Belin et al., 2022). To detect mild stages of KC with higher accuracy, a series of corneal topography technologies, for instance, Placido disc, photo-keratoscopy, keratometry, and computer-assisted video-keratography (Bevara & Vaddavalli, 2023; De Sanctis et al., 2007; Gordon-Shaag et al., 2012) have been widely used. Scheimpflug imaging, optical coherence tomography, and slit-scan tomography have also been used for KC early detection (Buzzonetti et al., 2020). The diagnostic methodologies discussed thus far necessitate using advanced clinical devices for data collection. These instruments, however, have limitations in that they could be more convenient and may require a skilled technician for operation and maintenance. Furthermore, for an accurate and reliable Keratoconus (KC) diagnosis, ophthalmologists and optometrists should concurrently consider various other clinical indicators of the disease (Agrawal, 2011).

Utilizing advancements in machine learning, image processing technologies, and smartphones' ubiquity, we can now develop accurate and easy-to-use imaging and diagnosis methods. For example, the brilliant KC consists of a 3D-printed Placido disc attachment, an LED light strip, and an intelligent app capturing the reflection of Placido rings. It utilizes an image processing pipeline to analyze the corneal image and applies the smartphone's camera parameters, Placido rings' 3D location, and pixel location of reflected rings for the KC diagnosis (Gairola et al., 2021). Another novel feature of our approach is capturing side photos of the eyes, processing these images to analyze corneal curvature at various angles, and identifying keratoconus based on the differences observed. This method is adaptable to multiple smartphone types with high detection accuracies for severe to moderate keratoconus (Askarian et al., 2018).

This manuscript describes a portable and easy-to-use methodology for diagnosing Keratoconus (KC). Our method leverages smartphones' versatility for detecting KC and pinpointing the corneal region impacted by the disease. As illustrated in Figure 1b, our procedure does not require a 3D-printed attachment; instead, we display a Placido disc on the smartphone's screen, directed towards the patient's eye, and capture images of the discs reflected from the cornea. This novel approach aids in not only identifying KC but also localizing the affected area of the cornea. Fig 1c provides a broader view of our principle where the Placido disc is displayed on the smartphone's screen. The smartphone's camera then captures images of the reflected discs from the right or left eye. These captured images are subsequently used for analysis and diagnosis. Furthermore, we have enhanced the Placido disc's design, as shown in Fig 1d-i, where the center ring has been transformed into a solid circle to facilitate graphic localization on the cornea. Fig 1d-ii exhibits images obtained during our experiments; the left side represents a non-KC cornea, while the right represents a cornea affected by KC.

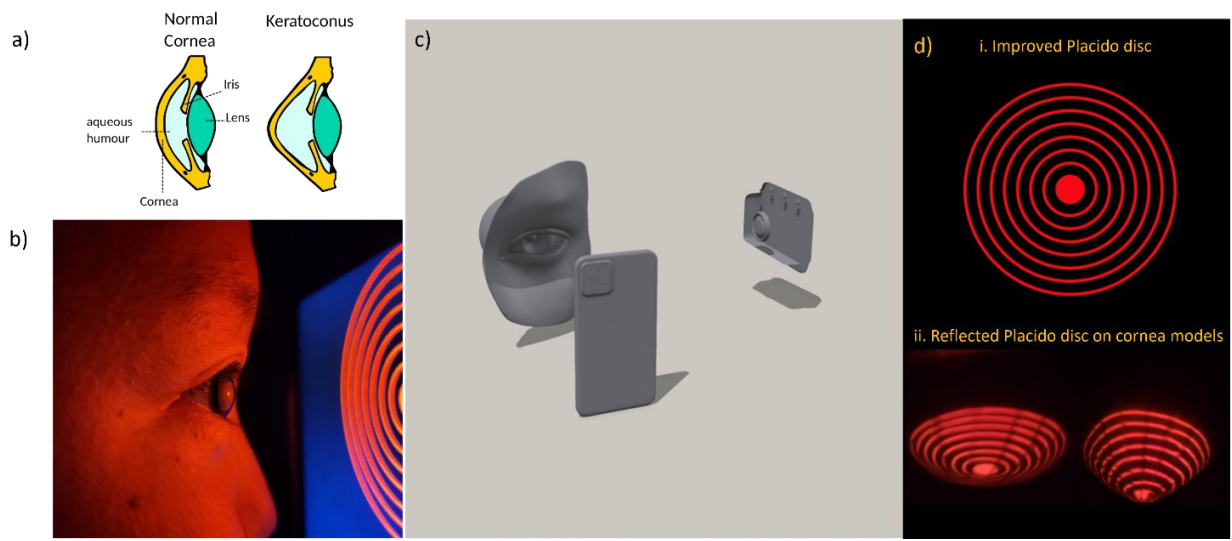

**Figure 1. Overview of portable Keratoconus diagnosis. a)** Demonstration of the non-KC eye and KC eye with the cone-shaped cornea. **b)** A rendering for our proposed approach, where the cornea reflected Placido rings. **c)** A broader view of our principle is where the Placido disc is displayed on the smartphone's screen, and the camera captures images from a side perspective. **d)** i. Improved Placido disc and ii. We reflected Placido disc on cornea models.

## 2. Methodology

### 2.1. Data preparation and preprocessing

In this study, we utilized a corneal disease eye model to simulate Keratoconus (KC), owing to the insufficiency of human data. The model was a GPI Anatomical Cornea Eye # 2780 (5x Life Size) (Cornea Eye – Model #2780). To facilitate the collection of images, we conceptualized and implemented a positioning apparatus for the corneas and the light source, as depicted in Figure 2c. As the model's thickness, at approximately 4 millimeters, substantially exceeds the typical human cornea thickness of 0.5 to 0.6 millimeters (Reinstein et al., 2008) by a factor of 6-8 times, it introduces a significant variable. The effect of this discrepancy is observed in the quality of the images collected. Specifically, the internal layer of the model causes a reflective interference with the Placido disc and generates distorted graphics. To mitigate the distortions attributable to the thickness of the models, we adhered reflection enhancement masks onto the external layers of both cornea models. We also designed an 8-circled Placido disc image that improves from the Placido discs of a keratoscope used in the optical clinic, as shown in Figure 1d-i. Our methodology involved capturing images from the apex of the apparatus and maintaining a perpendicular orientation between the camera and the smartphone, providing the light source. Hence, it is challenging to extract the whole circle in our collected images. Thus, we made the center circle as shown in Figure 1d i., the solid round located at the center of the Placido disc to assist in locating the center of reflection graphics. For the image collection, we used a Samsung Galaxy Z Fold3 smartphone equipped with cameras that have 12 MP sensors with either f/1.8 standard aperture lenses.

Firstly, we converted the color image to a binary image to reflect Placido disc extraction. We labeled the collected images as shown in Figure 3a and utilized a robust method to separate the Placido graphic and background with a single classification tree. We considered several features to train the classification tree, including pixel values in the HSV (Hue, Saturation, and Value) color space, components from the normalized RGB (Red, Green, Blue) color space, and chroma values as shown in Eq. 2.

$$[pixel_H, pixel_S, pixel_V, \frac{R}{R+B+G}, \frac{G}{R+B+G}, chroma] \qquad (2)$$

where $pixel_H$, $pixel_S$, and $picel_V$ represent the pixel values in HSV color space (Sobottka & Pitas, 1996), *R, G, B* values are the pixel values at R, G, B channels of the images. Then, we retained the pixels predicted at the Placido graphic by the classification tree and generated a binary image.

Secondly, we applied an open-close operation (Soille, 1999) to denoise. Variance in the surface curvature of the cornea model can result in uneven light reflection (Urone & Hinrichs, 2016). Specifically, areas with more significant curvature tend to capture more light, appearing brighter in images (Healey & Kondepudy, 1994). Furthermore, the smartphone, which displays the 8-circle Placido disc, is the light source. As a result, areas closer to this light source also exhibit increased brightness.

Additionally, inherent noise in the camera sensor can contribute to disparities in the captured image. These factors collectively lead to color distortions in the original photos, which can interfere with binarization. To minimize the impact of color distortions, we applied an opening-closing operation (Soille, 1999). It is a morphological operation to reduce noise and enhance the object of interest. The opening operation is an erosion operation followed by dilation. It helps to remove small white objects not belonging to Placido circles in the binary image and smooth the boundary of the larger objects (Urone & Hinrichs, 2016). The closing operation is dilation followed by erosion. It helps to fill small holes and connect slightly disconnected objects.

Thirdly, we corrected the orientation (Lee et al., 2017) of the denoised image for further analysis. The processed binary image visualized the enhanced Placido rings, which included a central ellipse and a series of arcs. We located all the connected components in the image and found the central ellipse by identifying the element with white pixels in the central bottom area. The orientation, determined by the angle between the x-axis and the ellipse's central axis, was used to rotate the entire image and perform further analysis.

The final step is to focus on the region of interest by removing the irrelevant parts of the image. The bounding boxes (Lempitsky et al., 2009) of the ellipse and arcs define a rectangle for cropping.

### 2.2. *k*-means clustering

K-means clustering is an unsupervised algorithm that can partition a given dataset into distinct clusters based on similarity (Hartigan & Wong, 1979). It is suitable for datasets with lower dimensionality and relatively robust to noise and outliers. With the results of Eq. 1 and the observation of collected images, we noticed notable differences in the dimensions of the captured images between the KC group and the non-KC group. Remarkably, the height differences were quite significant. Moreover, the images from the non-KC group tended to exhibit greater widths, indicating divergence between the two cohorts.

Thus, we extracted width and height features from cropped images, obtained two centroids from 2-dimensional features with k-means clustering, and implemented the k-means++ algorithm (Arthur & Vassilvitskii, 2007) to optimize initial values $c_1$ and $c_2$ with a weighted probability distribution, as shown in Eq. 3.

$$P = \frac{d^2(x_m, c_1)}{\sum_{j=1}^{n} d^2(x_j, c_1)} \tag{3}$$

where $c_1$ is the first centroid which is uniformly selected at random from the samples, $P$ is the probability of the other samples ($x_m$) are chosen to be $c_2$, $d(x, c_1)$ represents the distance between $c_1$ and the other samples, $x_j$ represents each sample in 2-dimension features, and $n$ represents the number of samples.

### 2.3. Logistic regression

Logistic regression allows the binary outcome model with one or more explanatory variables (Scott Long, 1997), evaluating the relationship between the categorical dependent variable and one or more independent variables through probability scores. In this case, the binary outcome is whether a cornea is affected by Keratoconus or not, and the independent variable is the inter-disc distance obtained from the reflected Placido disc graphic. The 'S' curve formed by the logistic function can be presented in Eq. 4.

$$P = \frac{1}{1 + e^{-(\beta_0 + \beta_1 x)}} \tag{4}$$

where $P$ represents the probability that a cornea is affected by KC, $x$ represents the inter-disc distances in distance matrixes as shown in Figure 4a, and 4b, $\beta_0$ and $\beta_1$ are coefficients learned from the dataset.

### 2.4. T-test

T-tests are commonly used to determine if there is a significant difference between the means of two groups. To demonstrate that the inter-disc distance is significantly different between KC and non-KC corneas, we proposed a null hypothesis (H0) that there is no significant difference in the inter-disc distances between the KC and non-KC groups.

Then, we calculated the t-score, which measures how far away the estimated parameter value is from the hypothesized value using Eq. 5.

$$t = \frac{m_1 - m_2}{\sqrt{\frac{\sigma_1^2}{n_1} + \frac{\sigma_2^2}{n_2}}} \tag{5}$$

where $m_1$ and $m_2$ are mean values of the two groups, $\sigma_1$ and $\sigma_2$ are their standard deviations, and $n_1$ and $n_2$ are a sample size of the two groups. A larger absolute t-score suggests a more significant difference between the groups. Then, the p-value is derived from the t-distribution table based on the t-score; a lower p-value indicates evidence against the null hypothesis. The effect size is also calculated based on Eq. 6.

$$d = \frac{t}{\sqrt{n_1 + n_2}} \tag{6}$$

Generally, a *d* value of 1 represents a difference of 1 standard deviation between groups, being considered a large effect size, which indicates that the two groups are substantially different.

### 2.5. Cohen's d

Cohen's d quantifies the effect size as a standardized difference between two means with raw data without a t-test (Cohen, 1992). It tells how substantially the two groups differ by calculating the difference between the two groups' means and normalizing it by the pooled standard deviation. We calculated Cohen's d following Eq. 7.

$$d = \frac{m_1 - m_2}{\sqrt{\frac{(n_1 - 1)\sigma_1^2 + (n_2 - 1)\sigma_2^2}{n_1 + n_2 - 2}}} \tag{7}$$

Roman 10-point bold in sentence structure (Sub-Section Heading Style). At most, there are three tiers of sections.

## 3. Results

### 3.1. Image preprocessing

In this paper, we applied a robust approach to extract reflected Placido discs as the diagram shown in Figure 3a. In the first step, we extracted the reflected graphic. Using a single classification tree, we generated a binary mask from the original photo based on the pixel value (Breiman, 2017). For testing the performance of the classification tree, 7 out of 25 images in each group are chosen randomly as a testing set. The single classification tree achieved 97.62% in the

KC cornea group and 95.23% in the non-KC cornea group accuracy of classifying our target graphic and background. Secondly, we implemented the "open-close" method to the binary mask, where the opening removed small white objects not belonging to Placido circles in the binary image and smoothed the boundary of the larger objects. The closing operation filled small holes and connected slightly disconnected objects.

Additionally, we found all connected components (He et al., 2009) in the binary image. We located the solid ellipse from the solid center ring by checking if the white pixels were at the center of the bounding box's bottom edge. Then, we identified the oval's long axis with image moments analysis technology and made the long axis align with the x-axis. In our last step, we retained the ROI (region of interest) for further analysis.

### 3.2. Two-stage KC diagnosis – Classification

The extracted reflected Placido discs are shown in Figure 3b, where the left image is from the Keratoconus-affected eye model, and the right image is from the non-KC eye model. The two images are on the same scale. This indicates the height of the cropped reflected Placido image, which is an essential feature for recognizing the KC cornea. It's also proven by Eq 1. that when KC develops, the distance between any of the two discs will increase. Hence, we proposed a two-stage KC diagnosis method: classification and localization stages.

In the classification stage, we applied *k*-means clustering to analyze and predict if the cornea model has KC or not, as shown in Figure 3c, where the blue points represent non-KC samples, red points represent KC samples, the yellow area represents a non-KC cluster, and green area represents KC cluster. The clear separation between clusters suggests significant, measurable differences between non-KC individuals and those with Keratoconus. The centroids, marked with an 'X' and centrally located within their respective clusters, also suggest that the extracted graphic of the Placido disc from the KC group shows a more significant height and smaller width.

Furthermore, the two clusters show some overlap along the width axis. Still, they are distinctly separated along the height axis, which indicates that the height changes are more indicative of the presence or progression of Keratoconus. Additionally, our implementation of the *k*-means clustering achieves an accuracy of 100% in classifying control and KC-affected corneas.

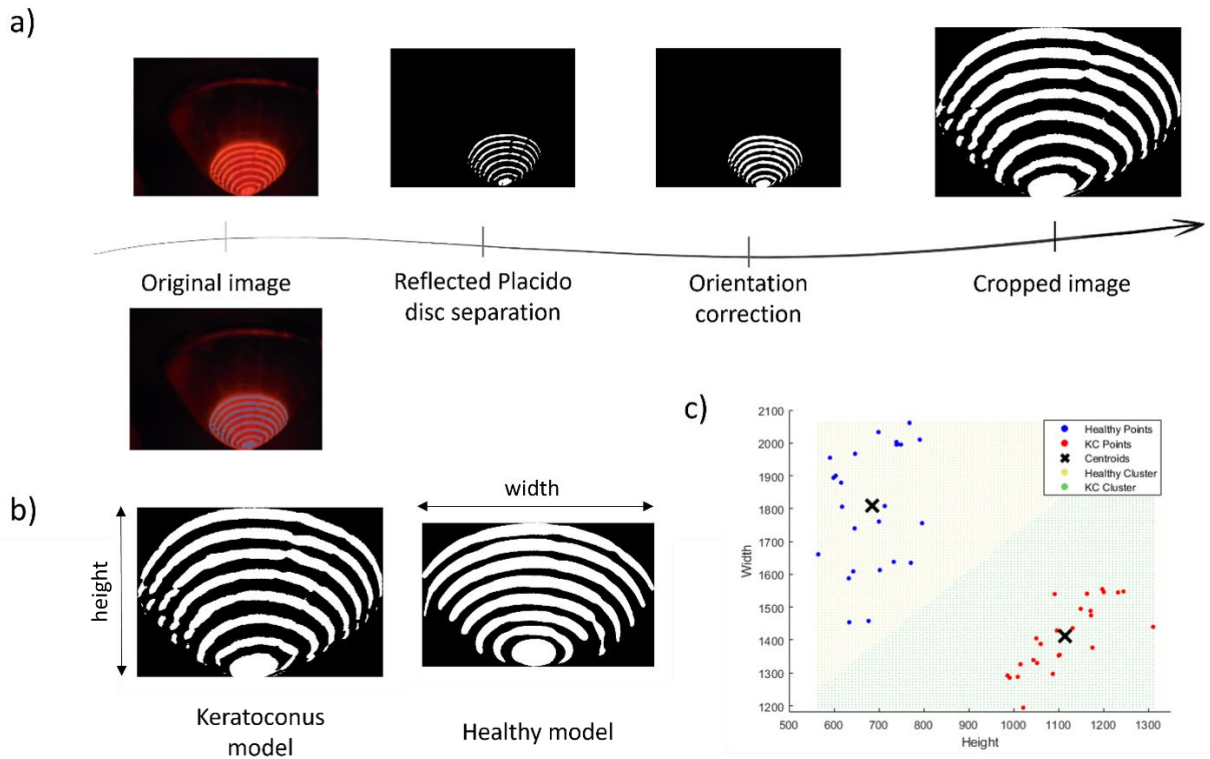

**Figure 3. Image preprocessing and KC classification. a)** Flowchart of image preprocessing including binary mask generation, orientation correction, and irrelevant parts removal **b)** Extracted reflected Placido discs **c)** Classification of Keratoconus and Non-KC Corneal Groups Using *K*-means Clustering.

### 3.3. Two-stage KC diagnosis – Localization

Keratoconus causes a localized thinning and protrusion of the cornea, which alters its topography. KC does not appear at a fixed location but manifests anywhere on the cornea. Thus, we visualize a distance matrix to pinpoint the exact location of the protrusion. This distance map helps in more accurate diagnosis and treatment planning, such as determining suitable sites for corneal cross-linking or intracorneal ring segments (ICRS) placement as performed conventionally in Alió's work (Alió et al., 2006).

Our first step is constructing a distance matrix, generating rays emanating from the ellipse's geometric center and extending outwards in an angular range between 45 and 135 degrees as shown in Figure 4a, the red points represent the intersections of the rays and rings. When the rays progressed from the center to the cornea's periphery, we implemented a color transition point detection algorithm on the extracted graphic as shown in Figures 4a and 4b, where 4a is from the non-KC cornea model and 4b is from the KC model. It helps to identify precise locations where there was a transition in color, which are red points shown in 4a and 4b. Then, we calculate the Euclidean distance between the center point of neighbor discs in the angular range to construct distance matrixes.

Secondly, we analyzed all the distances in distance matrixes and displayed the box plot in Fig 4c. The box plot demonstrates a clear difference in the distributions of the KC group and the Non-KC group: the median distance in the KC group, approximately 148.93, is significantly higher than the median of the Non-KC group, around 82.97. It indicates that the inter-disc distances are typically more significant in the KC group compared with the Non-KC group.

In our third step, we implemented logistic regression to explore the differences in inter-disc distances between the KC and non-KC groups. Figure 4d shows the blue points representing the non-KC group and the red points representing the KC group. The logistic regression 'S' curve indicates a shift towards higher distance values as the probability of being diagnosed with KC increases. This shift is evidence of different distances between the KC and non-KC corneas. The data distribution also suggests this difference - the two groups of points show a slight overlap. When the probability threshold is set at 0.5, the decision boundary, it is evident that only a few points fall within this region of maximum uncertainty. This further corroborates the idea that distance is essential in distinguishing KC and non-KC groups.

From our statistical analysis, as shown in Table 1, where the null hypothesis is that there is no difference in distance between KC and non-KC groups, there is evidence against the null hypothesis in the t-test. It underscores that the observed differences in distances between KC and non-KC corneas are highly unlikely to have occurred by chance. The absolute value of effect size observed from the *t*-test is 1.9015, which is a large effect size, emphasizing the stark contrast between the two groups. Similarly, Cohen's d (Lakens, 2013) achieves the value of 3.8030, implying that the mean of the KC group is approximately 3.8 standard deviations away from the mean of the non-KC group. It indicates a considerable difference in distances between the two groups. Lastly, our logistic regression model demonstrated an accuracy of 96.94%, proving inter-disc distance can effectively classify KC and non-KC conditions.

|           | t-score  | p value      | Effect size |
|-----------|----------|--------------|-------------|
| T-test    | 135.7408 | $<10^{-6}$   | 1.9015      |
| Cohen's d |          | 3.8030       |             |

| | |
|---|---|
| Logistic regression accuracy | 96.94% |

<div align="center">**Table 1. Statistical analysis for inter-disc distances**</div>

In the final step of the analysis, we implemented a color scheme to enhance the interpretability of our distance matrix. This visual enhancement more effectively differentiates the levels of corneal protrusion, with individual colors indicating diverse depths. It allows us to distinguish non-KC corneas from KC corneas and localizes the exact location of KC on the cornea.

In the final step of our analysis, we augmented the interpretability of our distance map by introducing a color map that reflects the severity of corneal protrusion, as shown in Figures 4e and 4f, where Figure 4e is generated based on the distance matrix of corresponding non-KC cornea model, and 4f is generated based on distance matrix of corresponding KC cornea model. Spanning an angular range from 45 to 135 degrees, this visual improvement distinguished varying inter-disc distances effectively, with the color spectrum representing these distances. More blueish colors, such as blues and greens, were assigned to shorter distances, corresponding to healthier cornea states, while warmer hues, including oranges and reds, depicted longer distances caused by protrusion, indicating more severe KC conditions. As the colors transition from cool to warm, they suggest a progression from non-KC status to KC-affected status, with the red tones indicating the most severe cases. The color map differentiates between non-KC and KC-affected corneas and precisely localizes the KC on the cornea, making it a valuable tool for accurate diagnosis and targeted treatment planning.

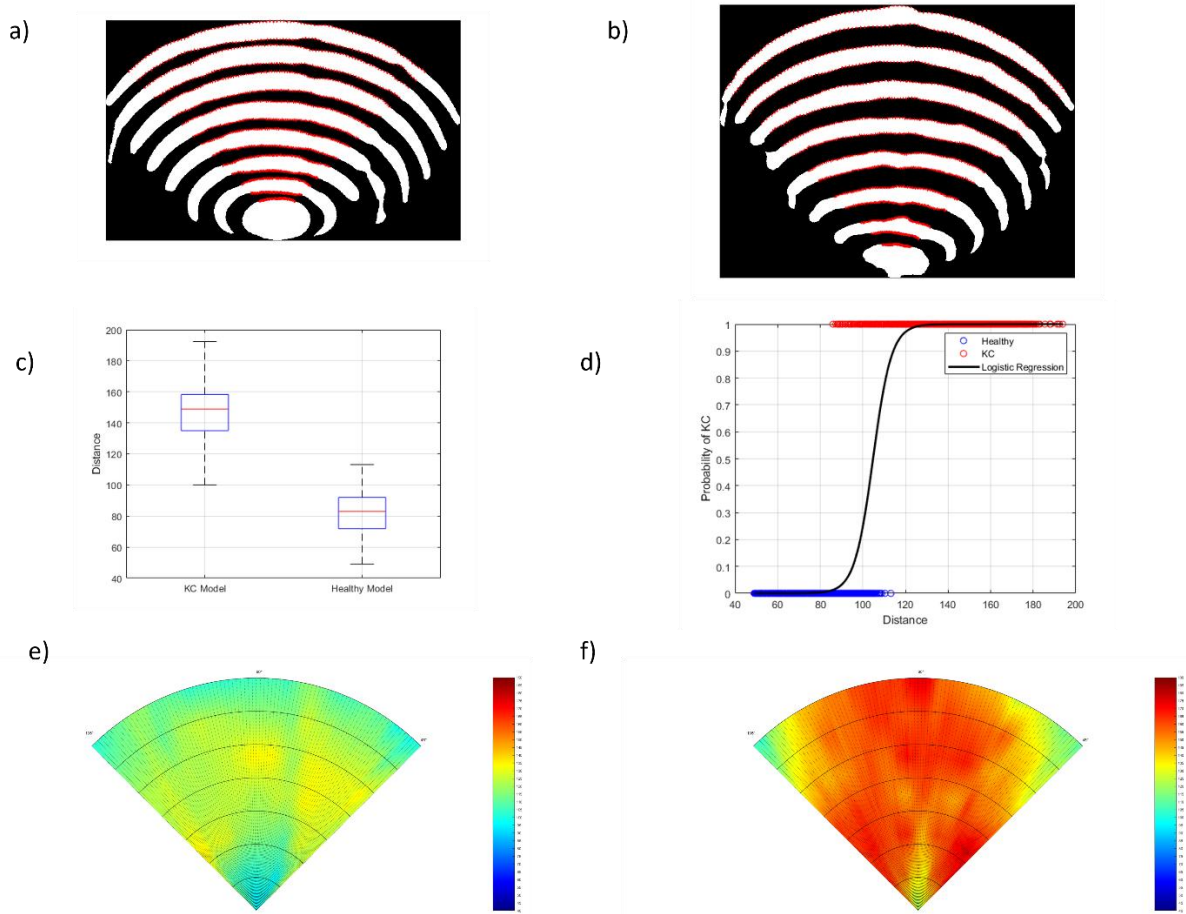

**Figure 4. Visualized distance matrix development and statistical analysis. a)** Distance matrix of a non-KC cornea model developed with inter-disc distances. **b)** The distance matrix of a KC cornea model was developed with inter-disc distances. **c)** Boxplot of inter-disc distances for KC model and non-KC model. **d)** Logistic regression based on inter-disc distances. **e)** Color map based on distance matrix of corresponding non-KC cornea model. **f)** The color map is based on the distance matrix of the corresponding KC cornea model.

## 4. Conclusion

Placido disc technology and its computer analysis have long been an established method for diagnosing keratoconus. However, it requires operator experience and costs a lot. This manuscript introduces a novel application of Placido disc technology using a smartphone platform. We utilize the smartphone camera to capture distinctive corneal images, allowing for an entirely fresh data extraction and analysis approach.

In our approach, we modified the traditional Placido disc by replacing the empty central ring with a solid circle, significantly improving the accuracy and consistency of the extracted reflection. As shown in Figure 3, compared to the empty ring design, the solid circle allows for the entire reflected shape to form a well-defined ellipse. This complete reflection enables precise orientation correction by detecting the long axis of the reflected shape and rotating it accordingly. With a consistently aligned reflection, subsequent feature extraction becomes more reliable. Specifically, for distance matrix generation, the rays used to compute inter-disc distances will always originate from a fixed starting point which is the center of the solid reflected ellipse, minimizing the bias in distance measurements compared to the fragmented reflections from a traditional Placido disc that is hard to localize center. Furthermore, when measuring the height and width of the extracted Placido disc reflection, the corrected orientation ensures more accurate dimension estimation, reducing errors caused by misalignment. This refinement ultimately enhances the overall precision of keratoconus classification and localization.

The crux of our diagnostic process is a two-tiered approach comprising a classification and localization stage. For the initial classification stage, we turned to k-means clustering, applying it to images of cropped reflected Placido discs. The results demonstrated a marked difference between non-KC corneas and those affected by Keratoconus.

Moving to the localization stage, we rolled out a distance matrix of the cornea, focusing on the analysis of inter-disc distances. Our deep-dive analysis indicated that Keratoconus severity significantly impacts these distances. To solidify this point, we employed a logistic regression model that underscored the statistically significant differences in inter-disc distances. This allowed us to distinguish between non-KC and KC-affected corneas with a remarkable classification accuracy of 96.94%. These findings were underpinned by a highly significant p-value of less than $10^{-6}$, an effect size of -1.9015, and a Cohen's d value of 3.8030.

Our comprehensive analysis not only helped distinguish non-KC corneas from those suffering from Keratoconus but also provided the location of the KC. A noteworthy addition was a color map in the distance matrix, improving the precise localization of KC on the cornea and boosting diagnostic accuracy. This inventive technique can refine treatment strategies by offering a lucid visualization of the affected areas. Because of widely used smartphones, our approach democratizes access to advanced diagnostic tools, making them available to a broader audience, including those in remote or underserved areas. It allows for rapid and convenient preliminary screening for keratoconus without requiring specialized equipment or trained personnel. In future work, we plan to conduct comprehensive clinical trials to assess the efficacy and accuracy of our diagnostic method. This will include gathering data from patients with different severities of keratoconus and exploring integrating additional diagnostic modalities, such as corneal biomechanics and tomography, ensuring our model is robust and reliable for real-world applications.